\documentclass[axodraw,prd,showpacs,epsf,graphicx,preprintnumbers]{revtex4}
\usepackage{amssymb}
\usepackage{graphicx}
\usepackage{dcolumn}
\usepackage{bm}
\usepackage[cp1251]{inputenc}
\usepackage[T2A]{fontenc}
\usepackage[russian]{babel}
\usepackage{amsmath}
\usepackage{feyn}
\usepackage{feynmf}

\newcommand{\la}{\lambda}
\newcommand{\lb}{\label}
\newcommand{\om}{\omega}
\newcommand{\al}{\alpha}

\newcommand{\vak}{\varkappa}
\newcommand{\pa}{{\partial}}
\newcommand{\ga}{\gamma}
\newcommand{\si}{\sigma}
\newcommand{\Si}{\Sigma}

\newcommand{\fr}{\frac}

\newcommand{\bw}{\begin{widetext}}
\newcommand{\ew}{\end{widetext}}
\newcommand{\be}{\begin{equation}}
\newcommand{\ee}{\end{equation}}
\newcommand{\ba}{\begin{eqnarray}}
\newcommand{\ea}{\end{eqnarray}}

\newcommand{\ep}{\epsilon}

\def\cM{{\cal M}}
\def\cV{{\cal V}}

\def\r{{\bf r}}

\def\k{{\bf k}}

\def\e{{\rm e}}

\begin{document}
\large
\title{Nambu-Goldstone explosion under brane perforation}
\author{D.V. Gal'tsov\footnote{E-mail: galtsov@physics.msu.ru}, E.Yu. Melkumova and S.
Zamani-Mogaddam}
\address{ Department of Physics, Moscow State University,119899,
Moscow, Russia}  
\begin{abstract}
We show that perforation of the three-brane  by  mass impinging upon
it from the five-dimensional bulk excites  Nambu-Goldstone spherical
wave propagating outwards with the velocity of light.  It is
speculated that such an effect can give rise to ``unmotivated''
energy release events  in the brane-world cosmological models.
\end{abstract}
\maketitle
\section{Introduction}
Brane-world cosmological scenarios   \cite{defect,RS} leave open the
possibility of presence of  matter in the bulk, at least in the form
of black holes \cite{Rubakov:2001kp}. Interaction of bulk black
holes with the three-brane is important both for the fate of the
brane itself \cite{Flachi:2007ev}, as well as  for the fate  of
black holes presumably created at colliders in TeV-scale gravity
\cite{escape}. This problem was extensively studied recently using
the Dirac-Nambu-Goto (DNG) equation on different black hole
backgrounds \cite{BBHS} leading to variety of interesting
predictions. Here we would like to investigate collision between the
mass and the three-brane as genuine two-body problem, treating both
of them on equal footing. Clearly, relativistic two-body problem
with field mediated interaction  has no exact solutions already in
the case of two point particles in flat space. The standard approach
is  perturbation theory in terms of the coupling constant. We
therefore adopt the same strategy. In order to maximally simplify
the problem, we will treat the mass as point-like, gravity ---
within the Fierz-Pauli theory, and we restrict  by an iterative
solution in terms of the coupling constant $\vak,$ related to the
five-dimensional gravitational constant $G_5$ as $\vak^2=16\pi G_5$.

Gravitational constant in five dimensions has dimension of
length$^3$. Combining it with the particle mass $m$ (dimension of
length$^{-1}$) and the brane tension $\mu$ (dimension of
length$^{-4}$) we have two length parameters: $ l=\vak^{-2}
\mu^{-1}\,,\; r_g=\vak  \sqrt{m}\,, $ the first  corresponding to
the curvature radius of the bulk in the RS II setup, and  the second
--- to gravitational radius of the mass $m$. To keep contact with
the RS II model we have to consider distances small with respect to
$l$, while to justify the linearization of the metric for the mass
$m$ we have to consider distances large with respect to $r_g$. So to
apply the linearized theory to both objects we have to assume $ l\gg
r_g$, or $m\mu^2 \vak^6 \ll 1 \,.$ Clearly, in this approach we will
not be able to capture such essentially non-linear phenomena as
formation of a hole in the three brane bounded by two-brane, and an
associated Chamblin-Eardley instability \cite{ChamEard}. Treating
the mass as point-like, we exclude formation of a hole, but instead
we will be able to describe another interesting manifestation of the
brane perforation: an excitation of the quasi-free Nambu-Goldstone
(NG) field.

Our model of collision  is therefore quite simple: we treat bulk
gravity at the linearized level on Minkowski background, with both
the brane and the particle being described by geometrical actions.
Gravitational interaction between them   is repulsive in the
co-dimension one, and the force does not depend on  distance. When
the mass pierces the brane reappearing on the other side of it, the
sign of this force changes, and  the brane gets shaken. We show that
this shake gives rise to  spherical NG wave which then expands
within the brane with the velocity of light independently on further
particle motion. Though we do not discuss here secondary effects due
to interaction of the NG wave with matter fields on the brane
\cite{KuYo,Bu}, it is clear that this interaction will transfer
energy to the brane matter modes. Thus, for an observer living on
the brane the act of brane perforation  by a massive body  from the
bulk will look like a sudden explosive event not motivated by any
visible reasons.
\section{Set up}

Consider the  three-brane propagating in the five-dimensional
space-time $\left(\cM_5\,,\;g_{MN}\right)$, whose world-volume
$\cV_4$ is given by the embedding equations
 $   x^{M}=X^{M}(\sigma_\mu),\; M=0,1,2,3,4 , $
parameterized by arbitrary coordinates  $\sigma_\mu,  \;
    \mu=0,1,2,3$ on $\cV_4$.
The DNG equation reads
 \be\lb{em}
\partial_\mu\left(  X_\nu^N
g_{MN}\ga^{\mu\nu}\sqrt{-\ga}\right)=\fr12 g_{NP,M}X^N_\mu
X^P_\nu\ga^{\mu\nu}\sqrt{-\ga}\,,
 \ee
where $X_\mu^M=\pa X^M/\pa\sigma^\mu$ are the tangent vectors,
$\gamma^{\mu\nu}$ is the inverse induced metric on
$\cV_4,\;\ga^{\mu\nu} \ga_{\nu\la} =\delta^\mu_\la$,
 \be\label{ceq}
  \ga_{\mu\nu}=X_\mu^M  X_\nu^N g_{MN}{\big
 |}_{x=X}\,,
 \ee
and $\ga={\rm det} \ga_{\mu\nu}$.

The energy-momentum tensor of the brane is
 \be\lb{TEN}
T_b^{MN}=\mu\int X^M_\mu X^N_\nu
\ga^{\mu\nu}\;\frac{\delta^5\left(x-X
(\si)\right)}{\sqrt{-g}}\;\sqrt{-\ga}\;d^4\si\, .
 \ee
where $\mu$ is the brane tension,

Bulk gravitational field will be described within the linearized
theory expanding the metric as
 \be
g_{MN}=\eta_{MN}+\vak h_{MN}\,,
 \ee
with $\eta_{MN}={\rm diag}(1,-1,-1,-1,-1)$,  and the metric
deviation $h_{MN}$ being considered as the Minkowski tensor. The
linearized Einstein equation in the harmonic gauge $\pa_M h^{MN}=
\pa^N h/2,\;h=h^P_P$ reads:
 \be\lb{ge} \square_5 h_{MN}=\vak
\left(T_{MN}-\frac13\eta_{MN}T\right)\,,
 \ee
where $\square_5=-\pa_M\pa^M$ is the bulk D'Alembertian, and the
metric potentials satisfy the superposition principle being the sum
of the brane potentials and the mass potentials:
$h_{MN}=h^b_{MN}+h^m_{MN}$. For consistency, the respective
energy-momentum tensors  at the right hand side must be flat space
divergenceless $\pa_N T^{b,m}_{MN}=0$, so, building them, we should
not take into account gravitational.

In  zero order in $\vak$ the brane is assumed to be unexcited
$X^M_0=\Si^M_\mu \si^\mu\,,$
 where $\Si^M_\mu$ are four constant bulk Minkowski vectors which will be
 normalized as
$\Si^M_\mu\Si^N_\nu\eta_{MN}=\eta_{\mu\nu}\,.$
Obviously, this is a solution to the Eq. (\ref{em}) for $\vak=0$,
and the corresponding induced metric is four-dimensional Minkowski
 \be
T_b^{MN}=\mu\int \Si^M_\mu\Si^N_\nu
\eta^{\mu\nu}\delta^5\left(x-X_0(\si)\right)d^4\si\,,\quad
T=T^P_P\,.
 \ee
One can further indentify the coordinates $\si^\mu$ on the
unperturbed brane with the bilk coordinates $\si^\mu=x^\mu$, in
which case $\Si^M_\mu =\delta^M_\mu$. Denoting the fifth coordinate
as $z=x^4$ we obtain the solution of (\ref{ge}):
 \be\lb{brgr}
h^b_{MN}=\frac{\vak \mu}2\left(\Si_M^\mu\Si_{N\mu}
-\frac43\eta_{MN}\right)|z|=\frac{\vak \mu}6{\rm
diag}(-1,1,1,1,4)|z|\,.
 \ee
this is a potential linearly growing on both sides of the brane.
Comparing this with the RS II metric
 \be
ds_{RS}^2=\e^{-2k|z_{RS}|}\eta_{\mu\nu}dx^\mu
dx^\nu-dz_{RS}^2\,,\quad k=\frac{\vak^2\mu}{12}\,,
 \ee
at the distance $z$ from the brane  small compared with the
curvature radius of the AdS bulk, $kz\ll 1$, so that
 $\e^{-2k|z_{RS}|}\simeq 1-2k|z_{RS}|$ we see that they differ
by coordinate transformation. Indeed, the gauge for the RS solution
is non-harmonic. To pass to the harmonic gauge one has to transfrom
the fifth coordinate as
 \be\lb{rstr}
z_{RS}=z-2kz^2{\rm sign }(z)\,.
 \ee
This reproduces our solution (\ref{brgr}). Note that this
transformation  is non-singular on the brane: $\pa z_{RS}/\pa z=1$
at $z=0$.

Consider now motion of the point mass in the bulk along the world
line $x^M(\tau)=(t(\tau),0,0,0,z(\tau))$:
 \be\lb{hit}
 \frac{d}{d\tau}\left({\dot x}^N g_{MN}\right)=
 \frac12 g_{PQ,M}{\dot x}^P{\dot x}^Q\,.
 \ee
We assume the mass moves in the positive $z$ direction and hits the
brane at $t=0$ having the velocity $v$, and correspondingly, ${\dot
t}=\ga,\;{\dot z}=\ga v,\; \ga=1/\sqrt{1-v^2} $. Then we find from
(\ref{hit}) that just before and after  collision
 \be\lb{acce}
{\ddot t}=\frac{\vak^2\mu}{6}\ga^2v\;{\rm sign }(z)\,,\quad {\ddot
z}=\frac{\vak^2\mu}{12}\ga^2(1+4v^2)\;{\rm sign }(z)\,.
 \ee
Particle energy and momentum $m\dot t,\, m\dot z$ have no
discontinuity at the location of the brane $z=0$, but their
derivative have. It can therefore  perforate  the brane without
loosing the energy-momentum, but with sudden change of acceleration.
The sign  in (\ref{acce}) corresponds to gravitational repulsion, as
could be expected in the co-dimension one case \cite{geod}.

Now compute the gravitational potentials of the mass constructing
the source energy-momentum tensor in zero order in $\vak$. With
gravitational interaction being neglected, the mass moves freely
with the velocity $v$, so
 \be x^M(\tau)=u^M\tau=\ga(1,0,0,0,v)\tau\,,\quad u^M=\rm{const.}
 \ee
Substituting the corresponding energy-momentum tensor
 \be
T^{MN}_m=m\int u^M u^N \delta(x-x(\tau)) d\tau\,.
  \ee
into the Eq. (\ref{ge}), we obtain
 \be\lb{hpart}
   h^{MN}_m =
-\fr{\vak m }{(2\pi)^2} \fr{\left(u_M
u_N-\fr13\eta_{MN}\right)}{\ga^2(z-v  t)^2+r^2 },
 \ee
where $r^2= (x^1)^2+(x^2)^2+(x^3)^2$. This is nothing but the
Lorentz-contracted five-dimensional Coulomb gravitational field of a
moving point particle.

\section{Brane deformation}
The transverse coordinates of the brane can be viewed as
Nambu-Goldstone bosons (branons) which appear as a result of
spontaneous breaking of the translational symmetry \cite{KuYo}.
These are coupled to gravity and matter on the brane via the induced
metric (for a recent discussion see \cite{Bu}). In our case of
co-dimension one there is one such branon, which is a massless field
coupled to bulk gravity.

To derive the NG wave equation we consider the deformation of the
brane $X^M=X^M_0+\delta X^M$ caused by gravitational field
$h^m_{MN}$ due to matter in the bulk, compute the induced metric and
linearize the resulting equation with respect to $\delta X^M.$ Only
transverse to the brane perturbation $\Phi(x^\mu)=\delta X^4$ is
physical, for which we obtain the wave equation
 \be\lb{NGEQ}
\square \Phi(x)=J(x),
  \ee
where $\square_4=-\pa_\mu\pa^\mu$ is the flat D'Alembertian on
$\cV_4$, and the source term $J=J^4$, where
 \be\lb{JN}
 J^N=\vak\left(\frac12 \pa^Nh_{PQ} -\pa_P h_Q^N\right)\Biggl\lvert_{z=0}
 \Si_\mu^P\Si_\nu^Q \eta^{\mu\nu},
 \ee
with $h_{PQ}=h_{PQ}^m$. Substituting here the potentials
(\ref{hpart}), we obtain explicitly:
 \be\lb{Jx}
J(x)=\frac{\la vt}{(r^2+\ga^2v^2t^2)^2}\,,\qquad
\la=\frac{m\vak^2\ga^2}{(2\pi)^2}\left(\ga^2 v^2+\frac13\right).
 \ee

The retarded solution of the Eq. (\ref{NGEQ}) reads:
 \be\lb{Fig}
\Phi=\int G_{\rm ret}(x-x') J(x')d^4x'=\frac1{(2\pi)^4}\int \frac
{\e^{-ik x  }}{k^2+2i\ep \om}\,J(k) \,d^4 k\,,
 \ee
where $k^\mu=(\om,\k)$ and $J(k)$ is the Fourier-transform of the
source:
 \be
J(k)=\int \e^{ik x  }J(x) d^4
x=\frac{2\pi^2\la}{\ga}\frac{i\om}{\om^2+\ga^2v^2\k^2}.
 \ee
Evaluating the integral we find the following solution consisting of
two terms
 \be\lb{solF}
\Phi=\frac{\la}{
2\ga^3}\,\left(\frac{F_0(r,t)}{r}-\frac{F_1(r,t)}{r}\right),
 \ee
where
 \be\lb{F}
F_0(r,t)=\frac{\pi}2\theta(t)\left[\ep(r+t)+\ep(r-t)\right]\,,\quad
F_1(r,t)= \arctan\frac{r}{\ga v t}. \ee The first part $F_0$,
proportional to  Heviside function of time $\theta(t)$, is zero
until the moment $t=0$ of perforation. It describes an expanding
wave caused by the perforation shake discussed above (the first term
in $F_0$ looks contracting, but actually it is a constant,
$\theta(t) \ep(r+t)=1$). This wave, propagating with the velocity of
light, does not correlate with further motion of the particle.

The second part  $F_1$ is non-zero and smooth both before ($t<0$)
and after ($t>0$) the perforation. It does correlate  with  position
of the mass  describing a continuous deformation of the brane caused
by its gravitational field. It is small when the particle is far
away from the brane, and grows up to the maximal absolute value
equal to $\pi/2$ when it approaches the brane.

 It is easy to verify
that for all $t\neq 0,\, r\neq 0$ this term satisfies the {\it
inhomogeneous} wave equation:
 \be\lb{arc0}
\square \;  \frac{F_1(r,t)}{r}
=\frac1{r}(\pa_t^2-\pa_r^2)F_1(r,t)=\frac{2\ga^3vt}{(r^2+\ga^2v^2t^2)^2},
 \ee
reproducing the right hand side of the Eq. (\ref{NGEQ}). But it has
finite and unequal limits as $t\to\pm 0$ for all $r\neq 0$:
 \be
\lim_{t\to\pm 0}\arctan\frac{r}{\ga v t}=\frac{\pi}2 \ep(t).
 \ee
As was already explained, change of sign is due to  change  of
direction of the gravitational force between the  brane and the
particle. The  repulsive nature of this force manifests itself in
signs: when the mass approach the brane ($t<0$)   the repulsive
deformation is directed along $z$, which corresponds to $\Phi>0$,
when the particle reappears on the other side of the brane it repels
the brane in the negative $z$ direction, thus $\Phi<0$. Since the
force does not vanish  and instantaneously changes sign at the
moment of perforation, the second part of the solution has singular
$t$-derivatives. Indeed, applying the box operator in the sense of
distributions we get an additional delta-derivative term at the
right hand side of the Eq. (\ref{arc0}) which was obtained for
$t\neq 0$:
 \be\lb{arc1} \square  \frac{F_1(r,t)}{r}=
 \frac{2\ga^3vt}{(r^2+\ga^2v^2t^2)^2}+\frac{\pi
\delta'(t)}{r}\,.
 \ee

The delta-derivative term describes the instantaneous shaking force
exerted  upon the brane. It excites the NG shock wave described by
the first term in (\ref{F}). Indeed, acting by D'Alembert operator
on the $F_0$ part of the solution, we find exactly the same
delta-derivative term:
 \be
\square \frac{F_0(r,t)}{r}=\frac{\pi \delta'(t)}{r}\,.
 \ee
Actually, the right-hand side arises as the second time derivative
of  Heviside function $\theta(t)$ entering $F_0$, while the
remaining factor $[\ep(r+t)+\ep(r-t)]/r$ describes spherical shock
waves satisfying the {\it homogeneous} wave equation:
 \be
\square \frac{\ep(r\pm t)}{r}=\frac1{r}\; (\pa_t^2-\pa_r^2)\ep(r\pm
t)=0\,.
 \ee
The sum of the two terms in (\ref{solF}) therefore has no
discontinuity at $t=0$, while the discontinuity at $t=r$ corresponds
to the expanding shock wave, satisfying the homogeneous wave
equation. One could expect that brane deformation caused by the
gravitational field of the mass could be tight to the mass motion,
i.e. be of the $F_1$ type only. However, as we have shown,
non-smoothness of gravitational interaction at the  moment of
piercing gives rise  to NG explosive wave $F_0$ which then
propagates freely along the brane.

Even more surprising is that in the limit of zero mass velocity our
solution remains non-zero:
 \be
\Phi_0=\lim_{v\to 0} \Phi=
\frac{m\vak^2}{48\pi}\,\frac{\ep(r-t)}{r}\,.
 \ee
Moreover, acting on this expression by box operator, one obtains
zero, except for the point $r=0$, at which  one has to perform
calculations in terms of distributions. By virtue of the identity
$\Delta \frac1{r}=-4\pi \delta^3(\r)$ we then find an extra term:
 \be\lb{Fist}
\square \Phi_0=Q_{B}\delta^3(\r)\,,\quad
Q_{B}=\frac{m\vak^2}{12}\ep(t)\,.
 \ee
This might seem paradoxical, since the source term (\ref{Jx}) in the
NG equation (\ref{NGEQ}) looks to be zero for $v=0$. The paradox is
solved if we regard the source $J(x)$ as distribution. It is easy to
see, that in the limiting cases $t\to 0$ or $v\to 0 ,\,$  $J(x)$
exhibits properties of the three-dimensional delta-function.
Denoting $\al=\ga v t$,  we have:
 \be
\lim_{\al\to \pm 0}
\frac{\al}{(r^2+\al^2)^2}=\Bigl\{\begin{array}{cc} 0,\;& {\rm
if}\;\;r\neq 0, \\ \pm\infty, &{\rm if}\;\;r=0.
\end{array}
 \ee
Since the box operator in (\ref{NGEQ}) contains the
three-dimensional Laplace operator, $\square=\Delta-\pa_t^2$, it is
reasonable to consider $J(x)$ in the sense of  distributions on
$R^3$. Now, the integral of $J(x)$ over the three-dimensional space
is $\al$-independent up to the sign and finite,
 \be
\int J(x) d^3 x=\frac{4\pi\la}{\ga} \int\limits_0^{\infty}
\frac{\al\,r^2}{(r^2+\al^2)^2} dr=\frac{\pi^2\la}{\ga}\ep(\al)\,,
 \ee
where $\ep(\al)=\al/|\al|$ so, taking into account that
$\ep(\al)=\ep(t)$, we obtain:
 \be \lim_{\al\to \pm 0}
J(x)=\frac{\pi^2\la}{\ga}\ep(t)\delta^3(\r)\,,
 \ee
which for $v=0$ reproduces the right hand side of (\ref{Fist}).
Remarkably, this limiting value is the same if we consider time in
the close vicinity of the perforation moment $t\to 0$ for any
velocity $v$ of the mass $m$, or if we consider the limit of small
velocity $v\to 0$. In the latter case of quasi-static perforation
this limit holds for any $t$, and since the coefficient $\la$
remains finite as $v\to 0$, the point-like source in  (\ref{NGEQ})
might be attributed to some  NG ``charge'' (shake charge).

This notion allows us to make distinction between two static
situations. The first is that of the point mass sitting on the brane
all the time. In this case, coming back to the Eq. (\ref{JN}) for
brane perturbations, we find that the source term at the right hand
side will be zero. This ''eternally'' siting on the brane mass is
not in fact the bulk particle. On the contrary, the bulk particle
arriving at the brane with zero velocity interacts gravitationally
with the brane in such a way that a non-zero NG charge as a source
term in the branon wave equation is produced. Therefore, it does
generate the NG wave (\ref{Fist}). This NG ``charge'' is a
manifestly non-conserved quantity, changing sign at the moment of
perforation. For an observer on the brane the perforation is felt as
a sudden shake, and the corresponding NG field will be not a static
Coulomb field, but an expanding wave.

\section{Conclusions}

We have considered a simple model of collision between an infinitely
thin three-brane and a point-like bulk particle interacting
gravitationally in five-dimensional space-time within the linearized
Einstein theory. In this setup the particle impinging normally on
the brane freely passes through it  to reappear on the other side.
Since the particle has no size, no hole is created in the
three-brane and
 no two-brane appears surrounding the hole. These phenomena which
have to arise in the case of  an impinging mass of finite radius (a
black hole) are beyond the scope of the linearized gravity theory we
adopt here. But as we have shown, perforation has another important
effect which {\it is} well captured by the linear gravity
approximation: detonation of a shock  NG expanding wave at the
moment of piercing. This might seem surprising, since no direct
non-gravitational force between the mass and the brane exists in our
model, and the particle   passes  through the brane feeling only its
gravitational field. Similarly, the brane is not hit by the mass in
the mechanical sense, but only reacts on its gravitational field.
But the  potential energy of the point mass in the field of the
brane immersed in space-time with co-dimension one is linearly
growing, so the force is repulsive and distance-independent. When
the mass pierces the brane, this finite repulsive force
instantaneously changes sign, so its time derivative behaves as
delta-function of time.

Apart from this shock NG wave expanding with the velocity of light,
the retarded solution contains the precursor/tail part which is non
zero both before and after the perforation. This component is in
direct correlation with  position of the particle, so its
measurement may serve as a tool to see invisible matter in the bulk.

Our model was inspired by the RS II setup. Indeed, using the
linearized Einstein theory we were able to reproduce the RS II
solution at distances small with respect the curvature radius of the
bulk. But differently from the RS approach, we have considered both
the brane and the particle on equal footing as test interacting
objects in the Minkowski background. This allowed us to tackle the
problem relativistically (in the special relativity sense) and to
reveal  existence of explosive NG wave triggered at the moment of
perforation. One can speculate that perforation of the three-brane
in the brane-world models may give rise to ``unmotivated'' explosive
events in the observed Universe. Indeed, the NG   field universally
interacts with matter on the brane via the induced metric
\cite{KuYo,Bu}, and, consequently, the NG explosion will transform
into the matter explosion. In absence of matter, gravitational
radiation will be excited, to see this it is enough to pass to the
second postlinear order of Einstein theory. NG explosion can be
expected to hold in the full non-linear treatment as well. Indeed,
excitation  of the brane oscillations in the black hole case is
likely to have been observed in numerical experiments
\cite{Flachi:2007ev}.

The work was  supported by the RFBR grant 08-02-01398-a.

\end{document}